\documentstyle[12pt,epsfig,amssymb]{article}
\def\spp{\Sigma_{\scriptscriptstyle ++}}
\def\spm{\Sigma_{\scriptscriptstyle +-}}
\def\smp{\Sigma_{\scriptscriptstyle -+}}

\def\Sp{S_{\scriptscriptstyle +}}
\def\Sm{S_{\scriptscriptstyle -}}
\def\Rp{R_{\scriptscriptstyle +}}
\def\Rm{R_{\scriptscriptstyle -}}
\def\zp{z^\prime}

\def\alps{\alpha_s}

\def\xip{\xi^\prime}
\def\zip{z^\prime}
\def\hsi{\hat{s}_1}
\def\hti{\hat{t}_1}

\def\delp{{\Delta^\prime}}

\def\msbar{${\rm{\overline{MS}}}$} 
\def\alps1{${\cal{O}}(\alpha_s^1)$}

\def\alpsq{${\cal{O}}(\alpha_s^2)$}
\def\Qt{{\tilde{Q}}}

\begin{document}
\setlength{\baselineskip}{0.75cm}
\setlength{\parskip}{0.45cm}
\begin{titlepage}
\begin{flushright}
DO-TH 98/14 \linebreak
August 1998
\end{flushright}
\vskip 0.8in
\begin{center}
{\Large\bf Heavy Quark Fragmentation in \\
Deep Inelastic Scattering}
\vskip 0.5in
{\large S.\ Kretzer and I.\ Schienbein}
\end{center}
\vskip .3in
\begin{center}
{\large Institut f\"{u}r Physik, Universit\"{a}t Dortmund \\
D-44221 Dortmund, Germany }
\end{center}
\vskip 0.5in
{\large{\underline{Abstract}}}

\noindent
We perform an analysis of semi-inclusive production of charm (D-mesons) in 
neutral current (NC) and charged current (CC) deep inelastic scattering (DIS)
at full \alps1. Our calculation is based on the heavy quark 
scheme developed by Aivazis, Collins, Olness and Tung (ACOT) where we
include an \alps1 calculation of quark scattering contributions for
general masses and couplings. 
We review the relevant massive formulae and subtraction terms and discuss 
their massless limits. We show how the charm 
fragmentation function can be measured in CC DIS and we 
investigate whether the charm production dynamics
may be tested in NC DIS. We also discuss finite initial state quark mass effects
in CC and NC DIS.
\end{titlepage}
%
\noindent

\section{Introduction}
In a recent paper \cite{kschie1} we have analyzed heavy quark initiated 
contributions to fully inclusive deep inelastic (DI) structure functions.
Towards lower values of the Bjorken variable $x$,
heavy (charm) quarks are produced in about 20 \% of the neutral current (NC)
\cite{h1,zeus} and charged current (CC) \cite{ccfrsf}
deep inelastic events in lepton-nucleon collisions. Therefore in this 
kinematical range, heavy quark events contribute an important component to 
the fully inclusive DI structure functions of the nucleon.  
However, due to acceptance losses
this component can usually not be measured directly by inclusively tagging on 
charm events and more differential observables have to be considered
like $E_D$ \cite{cdhsw,ccfrlo,ccfrnlo}, $p_T$ or $\eta$ \cite{h1,zeus} 
distributions, where $E_D$, $p_T$ and $\eta$ are
the energy, transverse momentum and pseudorapidity of the charmed hadron
produced, i.e.\ mainly of a $D^{(\ast)}$ meson. 
In this article we consider $E_D$ spectra represented by the usual 
scaling variable $z$ defined below. Within DIS 
the charmed hadron energy spectrum is the distribution which is most sensitive
to the charm fragmentation process and may give complementary information to
one hadron inclusive $e^+e^-$ annihilation which is usually chosen to define
fragmentation functions (FF's) \cite{aemp,nasweb}. 
A well understanding of charm fragmentation is
essential for any charm observable, e.g.\ the normalization of $p_T$ and $\eta$
distributions in photoproduction is substantially influenced by the hardness
of the FF \cite{cagre,zeusgamma}.
The $z$ distribution of charm 
fragments is directly measured in CC neutrinoproduction 
\cite{cdhsw,ccfrlo,ccfrnlo,gkr2}
and may give insight into details of the charm production dynamics in NC 
electroproduction. It has e.g.\ been shown in \cite{h1,zeus}, that the energy 
spectrum of $D^{(\ast)}$-mesons produced at the $eP$--collider 
HERA may be able to discriminate between
intrinsic and extrinsic production of charm quarks. 

Intrinsic heavy quark densities may arise due to a nonperturbative
component of the nucleon wave function \cite{incc}
or due to a perturbative resummation \cite{acot,thoro,mrrs,bmsn}
of large quasi-collinear logs [$\ln (Q^2/m^2)$; $Q$ and
$m$ being the virtuality of the mediated gauge boson and the 
heavy quark mass, respectively] arising at any order in fixed order extrinsic
production (or fixed order perturbation theory: FOPT). 
Here we will only consider the latter possibility for inducing
an intrinsic charm density $c(x,Q^2)$ which is concentrated at small $x$ and 
we will ignore nonperturbative components which are expected to be located at
large $x$ \cite{incc}. 
Technically the resummation of large perturbative logs proceeds
through calculating the boundary conditions for a transformation of the
factorization scheme \cite{ct,bmsn,melnas}, 
which is switched from $n_f$ to $n_f+1$ active,
massless flavors, canonically at $Q^2=m^2$. For fully inclusive DIS 
the Kinoshita-Lee-Nauenberg theorem confines all quasi-collinear logs 
to the initial 
state such that they may be absorbed into $c(x,Q^2)$.
For semi-inclusive DIS (SI DIS) also final state collinearities 
arise which are resummed in 
perturbative fragmentation functions $D_i^c(z,Q^2)$ 
(parton $i$ decaying into charm quark $c$)
along the lines of \cite{melnas,cagre}. The scale dependence of 
$c(x,Q^2)$ and  $D_i^c(z,Q^2)$ is governed by massless renormalization
group (RG) evolution.

Besides this {\it{zero mass variable flavor number scheme}}, where
mass effects are only taken care of by the boundary conditions for $c(x,Q^2)$
and $D_i^c(z,Q^2)$, variable flavor number schemes have been formulated 
\cite{acot,thoro,mrrs,bmsn}, which aim 
at resumming the quasi-collinear logs as outlined above
while also keeping power suppressed terms of
${\cal{O}} [(m^2/Q^2)^k]$ in the perturbative coefficient functions. Our reference
scheme for this type of schemes will be the one developed by  
Aivazis, Collins, Olness and Tung (ACOT) \cite{aot,acot}. 
In the ACOT scheme full dependence on the heavy quark mass is kept in graphs 
containing heavy quark lines. This gives rise to the above mentioned 
quasi-collinear 
logs and to the power suppressed terms. While the latter are regarded as mass 
corrections to the massless, dimensionally regularized, standard coefficient 
functions (e.g.\ in the 
${\overline{{\rm{MS}}}}$ scheme), the former are removed numerically by  
subtraction terms, which are obtained from the small mass limit of the
massive coefficient functions.

The outline of our paper will be the following: In section 2 we will shortly
overview the relevant formulae for SI DIS for general masses
and couplings including quark scattering (QS) and boson gluon fusion (GF)
contributions up to \alps1. 
We will thereby
present our ACOT based calculation for the $QS^{(1)}$\footnote{
Bracketed upper indices count powers of $\alpha_s$.
}
component of SI structure functions. 
In section 3 we will analyze the charm fragmentation function in CC and
NC DIS. In section 4 we draw our conclusions and
some uncomfortably long formulae are relegated to an appendix.

\section{Semi-Inclusive Heavy Quark Structure Functions}
\label{sec2}

This section presents the relevant formulae for one (heavy flavored) hadron
inclusive DIS structure functions. The contributions from scattering events
on massive quarks are given up to \alps1 in section \ref{scattq} and GF 
contributions are briefly recalled in section \ref{glufu}. Section \ref{subterms}
presents all subtraction terms which render the structure functions infrared
safe and includes a discussion of these terms.

\subsection{Scattering on Massive Quarks} \label{scattq}

We consider DIS of the virtual Boson $B^\ast$ with momentum $q$
on the quark $Q_1$ with mass $m_1$
and momentum $p_1$
producing the quark $Q_2$ with mass $m_2$ and momentum $p_2$.
The latter fragments into a heavy quark
flavored hadron $H_{Q_2}$, 
e.g.\ a $\left|Q_2 \bar{q}_l\right>$ meson, $q_l$ being any light
quark. Phenomenologically most prominent are of course charm quarks fragmenting
into $D^{(\ast)}$-mesons which are the lightest heavy flavored hadrons.

We will strictly take over the formulae and notations of 
our inclusive analysis in \cite{kschie1}
whenever possible and merely extend them for SI DIS considered here.
In particular we take over the definition of the structure functions ${\cal{F}}_i$
given in terms of the usual experimental structure functions $F_i$ in 
Eq.\ (7) in \cite{kschie1}, i.e.
\begin{eqnarray}
\label{def}
{\cal{F}}_1&=&\frac{2 \Delta}{\Sp \spp- 2 m_1 m_2 \Sm}\ F_1 \\ \nonumber
{\cal F}_2&=&\frac{2 Q^2}{\Sp \Delta}\ \frac{1}{2 x}\ F_2 \\ \nonumber
{\cal F}_3&=&\frac{1}{2 \Rp}\ F_3\ \ \ , 
\end{eqnarray}
with
\begin{equation}
\label{pmpm}
\Sigma_{\pm \pm}\equiv Q^2\pm m_2^2\pm m_1^2\ \ \ .
\end{equation}
In Eq.\ (\ref{def}) we use the shorthand 
$\Delta \equiv \Delta[m_1^2,m_2^2,-Q^2]$ 
, where the usual
triangle function is defined by 
\begin{equation}
\Delta[a,b,c]=\sqrt{a^2+b^2+c^2-2(a b+ b c + c a)}\ \ \ .
\end{equation}
The vector ($V$) and axial vector ($A$) couplings of the 
${\overline{Q}}_2 \gamma_\mu 
(V-A \gamma_5) Q_1$ quark current enter via the following combinations:
\begin{eqnarray}
\label{couplings}
S_{\pm}&=&V{V}^{\prime} \pm A {A}^{\prime}
\nonumber\\
R_{\pm}&=&(V A^{\prime}\pm V^{\prime} A)/2
\end{eqnarray}
where $V,A \equiv V^{\prime},A^{\prime}$ in the case of pure $B$ scattering
and $V,A \neq V^{\prime},A^{\prime}$ in the case of $B, B^{\prime}$ 
interference (e.g. $\gamma, Z^0$ interference in the standard model).

Since we want to investigate the energy spectrum of charm fragments,
we introduce the Lorentz-invariant $z\equiv p_{H_{Q_2}}\cdot p_N/q\cdot p_N$ 
which reduces to the energy $E_{H_{Q_2}}$ scaled to its maximal value 
$\nu=q_0$ in the target rest
frame. Therefore in contrast to Eq. (A4) in \cite{kschie1} we do not integrate 
the tensor ${\hat{\omega}}^{\mu\nu}$ over the 
full partonic phase space but keep it differential in the corresponding
partonic variable $z^\prime \equiv p_2\cdot p_1/q\cdot p_1$
or the mass corrected variable ${\hat{z}}$ which is defined below. 
In order to obtain hadronic observables we have to extend the ansatz
of Eq.\ (17) in \cite{aot} such that it includes a nonperturbative hadronization
function $D_{Q_2}$. 
In the limit of vanishing masses the massless parton model expressions
have to be recovered. Our ansatz will be
\begin{equation}
\label{ansatz}
W^{\mu\nu} = \int\ \frac{d\xi}{\xi}\ \frac{d\zeta}{\zeta} 
\ Q_1(\xi,\mu^2)
\ D_{Q_2}(\zeta,[\mu^2])
\ {\hat{\omega}}^{\mu\nu}|_{\{p_1^+=\xi P^+; z=\zeta \hat{z}\}}\ \ \ ,
\end{equation}
where $\mu$ is the factorization scale,  
${\hat{z}}=z^\prime / z^\prime_{LO}$ with
$z^\prime_{LO}=\spp / \spm$
and $v^+\equiv (v^0+v^3)/\sqrt{2}$ for a general vector $v$.
$W^{\mu\nu}$ is the usual hadronic tensor and ${\hat{\omega}}^{\mu\nu}$ 
is its partonic analogue. 
Eq.\ (\ref{ansatz}) defines the fragmentation function $D_{Q_2}$ to be a 
multiplicative factor multiplying inclusive structure functions at LO/Born
accuracy, i.e.
\begin{equation}
{\cal F}_{i=1,2,3}^{QS^{(0)}} (x,z,Q^2) = 
{\cal F}_{i=1,2,3}^{QS^{(0)}} (x,Q^2)\ D_{Q_2}(z,[Q^2]) 
\ \ \ ,
\end{equation}
where the ${\cal F}_{i=1,2,3}^{QS^{(0)}}(x,Q^2)$ are defined and given in
\cite{kschie1}. The scale dependence of $D_{Q_2}$ is bracketed here and in the 
following because it is optional; a more detailed   
discussion on this point will be given at the end of section \ref{subterms}.
We do not construct our ansatz in Eq.\ (\ref{ansatz})
for the convolution of the fragmentation 
function along light front components for the outgoing particles which would
only be Lorentz-invariant for boosts along a specified axis. Since the final
state of DIS is spread over the entire solid angle it has no preferred axis as 
defined 
for the initial state by the beam direction of collider experiments.
Note that 
eq.\ (\ref{ansatz}) is in agreement with usual factorized 
expressions for massless initial state quanta, e.g.\ in \cite{kschie2}, since
there $m_1=0$ such that $z^\prime_{LO}=1$. 

Up to \alps1 the hadronic structure functions for scattering on a heavy quark read 
\begin{eqnarray} \label{QS1}
{\hat{{\cal{F}}}}_{i=1,2,3}^{QS^{(0+1)}} (x,z,Q^2,\mu^2)
&=& Q_1(\chi,\mu^2)\ D_{Q_2}(z,[\mu^2])
+\frac{\alpha_s(\mu^2)}{2 \pi}\int_{\chi}^1\frac{d\xip}{\xip}
\int_z^1\frac{d{\hat{z}}}{{\hat{z}}} \\ \nonumber
&\times&
\left[Q_1\left(\frac{\chi}{\xip},\mu^2\right)\ {\hat{H}}_i^q(\xip,\zp,\mu^2)
\ \Theta_q\  \right]
\ D_{Q_2}\left(\frac{z}{{\hat{z}}},[\mu^2]\right)\ \ ,
\end{eqnarray}
with \cite{aot,kschie1}
\begin{equation}
\label{chi}
\chi = \frac{x}{2 Q^2}\ \left(\ \spm\ +\ \Delta\ \right)\ \ \ . 
\end{equation}
In Eq.\ (\ref{QS1}) and throughout this paper we set the renormalization scale
equal to the factorization scale.
The kinematical boundaries of the phase space in Eq.\ (\ref{QS1}) 
are introduced by
the theta function cut $\Theta_{q}$. In the massless limit 
$\Theta_{q}\rightarrow 1$. The precise arguments of $\Theta_{q}$
are  set by the kinematical requirement
\begin{equation}
\label{boundary} 
\zp_{min} < \zp < \zp_{max}\ \ \ ,
\end{equation}
with
\begin{eqnarray} \nonumber
\zp_{max\atop min} = \frac{
\pm \Delta[{\hat{s}},m_1^2,-Q^2]({\hat{s}}-m_2^2)+(Q^2+m_1^2+{\hat{s}})
({\hat{s}}+m_2^2)
}
{
2 {\hat{s}}({\hat{s}}-m_1^2+Q^2)
}\ \ \ ,
\end{eqnarray}
where ${\hat{s}}=(p_1+q)^2$.
Note that Eq.\ (\ref{boundary}) also poses an implicit constraint on $\xip$
via $\zp_{min} < \zp_{max}$.

The ${\hat{H}}_i^q$ for nonzero masses are obtained in exactly the same way 
as outlined for fully inclusive structure functions in \cite{kschie1}
if the partonic phase space is not fully integrated over. 
They are given by
\begin{eqnarray} \nonumber
{\hat{H}}_i^q(\xip,\zp)&=&C_F\ \Big[
\ (S_i+V_i)\ \delta(1-\xip)\ \delta(1-{\hat{z}})  \\ \label{Hiq} 
&+& 
\frac{1-\xip}{(1-\xip)_{+}}
\ \frac{z^\prime_{LO}}{8}\ \frac{\hsi+\spm}{\Delta^\prime}
\ N_i^{-1}\ \hat{f}_i^Q(\hsi,\hti)\Big]
\end{eqnarray}
with the normalization factors
\begin{eqnarray}
N_1&=&\frac{\Sp \spp- 2m_1 m_2 \Sm}{2 \Delta},\qquad N_2=\frac{2 \Sp \Delta}{\Delta^\prime},\qquad
N_3=\frac{2 \Rp}{\Delta^\prime}\ ,
\nonumber
\end{eqnarray} 
and the mandelstam variables 
\begin{eqnarray*}
\hsi(\xip)&\equiv& (p_1+q)^2-m_2^2 \\
&=&\frac{1-\xip}{2 \xip}[(\Delta-\spm)\xip+\Delta+\spm]\ \\
\hti(\xip,\zp)&\equiv& (p_1-p_3)^2-m_1^2 \\
&=&[\hsi(\xip)+\spm](\zp-z_0^\prime) 
\end{eqnarray*}
where $\displaystyle z_0^\prime=\frac{\hsi+\spp}{\hsi+\spm}$ 
is the would-be pole of the ${\hat{t}}$-channel propagator
and we use $\Delta^\prime\equiv\Delta[{\hat{s}},m_1^2,-Q^2]$.
$S_i$ and $V_i$ are the soft real and virtual contribution to 
${\hat{H}}_i^q$, respectively.
They can be found in appendix C of \cite{kschie1} whereas the 
$\hat{f}_i^Q(\hsi,\hti)$ are listed in appendix A of this paper.
In the massless limit the ${\hat{H}}_i^q$
reduce to the \msbar\ coefficient functions in
\cite{aemp,fupe} up to some divergent subtraction terms which we will
specify in section \ref{subterms}.

\subsection{Gluon Fusion Contributions at \alps1} \label{glufu}

The semi-inclusive coefficient functions for GF production of massive
quarks have been obtained for general masses and couplings in \cite{kschie2}
and are only briefly reviewed here for completeness. They are given by
\begin{eqnarray}  \nonumber
{\hat{F}}_{1,3}^{GF}(x,z,Q^2,\mu^2) &=& \int_{ax}^{1} \frac{dx'}{x'} 
              \int_z^1
      \ \frac{d\zeta}{\zeta}\ g(x',\mu^2)\ f_{1,3}(\frac{x}{x'},\zeta,Q^2)
      \ D_{Q_2}(\frac{z}{\zeta},[\mu^2])\ \Theta_g  \\ \nonumber \\
{\hat{F}}_{2}^{GF}(x,z,Q^2,\mu^2) &=& \int_{ax}^{1} \frac{dx'}{x'} 
              \int_z^1
      \ \frac{d\zeta}{\zeta}\ x'g(x',\mu^2)\ f_{2}(\frac{x}{x'},\zeta,Q^2)
      \ D_{Q_2}(\frac{z}{\zeta},[\mu^2])\ \Theta_g
\label{GF}
\end{eqnarray}
\noindent
with $ax=[1+(m_1+m_2)^2/Q^2]x$.
The $\Theta_g$ cut guarantees that $\zeta_{min} < \zeta < \zeta_{max}$ where 
$\zeta_{min,max}$ are given in Eq.\ (3) of \cite{kschie2}. Similarly to
Eq.\ (\ref{boundary}) $\zeta_{min} < \zeta_{max}$ may also constrain the phase
space available for the $x^\prime$ integration.

\subsection{Subtraction Terms}
\label{subterms}

It requires three \msbar\ subtraction terms to render the double convolutions
in Eqs.\ (\ref{QS1}), (\ref{GF}) infrared safe: 
\begin{eqnarray} \nonumber
{\cal F}_i^{SUB_q} (x,z,Q^2,\mu^2)&=&\frac{\alpha_s(\mu^2)}{2 \pi}C_F 
\int_{\chi}^1\frac{d\xip}{\xip}
\left[\frac{1+{\xip}^2}{1-\xip}\left(\ln \frac{\mu^2}{m_1^2}-1
-2 \ln(1-\xip)\right)\right]_{+} \\ 
&\times&
Q_1\left(\frac{\chi}{\xip},\mu^2\right)\ D_{Q_2}(z,[\mu^2])
\label{SUBq}
\end{eqnarray}
\begin{equation}
\label{SUBg}
{\cal F}_i^{SUB_g} (x,z,Q^2,\mu^2)= D_{Q_2}(z,[\mu^2]) 
\ \frac{\alpha_s(\mu^2)}{2 \pi}
\ \ln \frac{\mu^2}{m_1^2} 
\ \int_{\chi}^1\frac{d\xip}{\xip}\ P_{qg}^{(0)}(\xip)   
\ g\left(\frac{\chi}{\xip},\mu^2\right)\ \ \ 
\end{equation}  
\begin{eqnarray} \nonumber
{\cal F}_i^{SUB_D} (x,z,Q^2,\mu^2)&=&\frac{\alpha_s(\mu^2)}{2 \pi}C_F 
\int_{z}^1\frac{d\zip}{\zip}
\left[\frac{1+{\zip}^2}{1-\zip}\left(\ln \frac{\mu^2}{m_2^2}-1
-2 \ln(1-\zip)\right)\right]_{+} \\ 
&\times&
D_{Q_2}\left(\frac{z}{\zip},\mu^2\right)\ Q_1\left(\chi,\mu^2\right)\ \ \ ,
\label{SUBD}
\end{eqnarray}
where $P_{qg}^{(0)}(\xip) = 1/2\ [{\xip}^2+(1-\xip)^2]$. 
Note that $SUB_g$ in Eq.\ (\ref{SUBg}) differs slightly
from Eq.\ (6) in \cite{kschie2} because we are allowing for
a nonzero initial state parton mass
$m_1$ here which we did not in \cite{kschie2}. 

The subtraction terms define the running of the initial state quark density
($SUB_q$, $SUB_g$) and the final state fragmentation function ($SUB_D$) in
the massless limit. They remove collinear logarithms and scheme defining
finite terms from the convolutions in  Eqs.\ (\ref{QS1}), (\ref{GF}) and  
they are constructed such that the massless 
\msbar\ results of \cite{aemp,fupe} are recovered in the limit 
\begin{equation}
\label{limit2}
\lim_{m_{1,2}\to 0} \left[{\hat{{\cal F}}}_i^{QS^{(0+1)}+GF} (x,z,Q^2,\mu^2)
-{\cal F}_i^{SUB_q+SUB_g+SUB_D} (x,z,Q^2,\mu^2)\right]=
{\cal F}_i^{(1),\overline{\rm{{MS}}}} (x,z,Q^2)\ \ \ ,
\end{equation}
where $SUB_q$ and $SUB_D$ regularize ${\hat{{\cal F}}}_i^{QS^{(0+1)}}$
whereas $SUB_g$ regularizes ${\hat{{\cal F}}}_i^{GF}$.
Contrary to the fully inclusive $SUB_g$ term in Eq.\ (16) of \cite{kschie1}
there is no need to include an additional $\sim \ln (\mu^2/m_2^2)$ subtraction
in Eq.\ (\ref{SUBg}) because the ${\zeta}^{-1}$ ${\hat{u}}$-channel singularity
of the massless limit of $GF^{(1)}$ is located at $\zeta=0$ and is outside the 
integration volume of Eq.\ (\ref{GF}).  
If only initial state subtractions, i.e.\ $SUB_q$ and $SUB_g$, are considered 
and final state subtractions, i.e.\ $SUB_D$, are not performed one
reproduces, in the limit $m_1\rightarrow 0$ the results in \cite{gkr2} for
producing a heavy quark from a light quark. Note that in this case the 
fragmentation function $D_{Q_2}$ should be taken scale-{\it{in}}dependent,
say of the Peterson form \cite{peterson}.
In the limit where also the final state quark mass $m_2$ approaches zero 
and the final state subtraction term
$SUB_D$ is subtracted from the results in \cite{gkr2} the massless
quark results in \cite{aemp,fupe} are obtained and a running of $D_{Q_2}$ is
induced via a RG resummation of final state collinear logs as formulated
for one hadron inclusive $e^+e^-$ annihilation in \cite{melnas,cagre}.

Apart from removing the long distance physics from the coefficient functions 
the subtraction terms set the boundary conditions for the intrinsic
heavy quark density  $Q_1$ \cite{ct}
and the perturbative part of the heavy quark
fragmentation function $D_{Q_2}$ \cite{melnas}: 
\begin{eqnarray} \label{boundq}
Q_1(x,Q_0^2)&=&
\frac{\alpha_s(Q_0^2)}{2 \pi}
\ \ln \frac{Q_0^2}{m_1^2} 
\ \int_{x}^1\frac{d\xi}{\xi}\ P_{qg}^{(0)}(\xi)   
\ g\left(\frac{x}{\xi},Q_0^2\right) \\ \nonumber
D_{Q_2}(z,\Qt_0^2)&=&
\frac{\alpha_s(\Qt_0^2)}{2 \pi}C_F 
\int_{z}^1\frac{d\zip}{\zip}
\left[\frac{1+{\zip}^2}{1-\zip}\left(\ln \frac{\Qt_0^2}{m_2^2}-1
-2 \ln(1-\zip)\right)\right]_{+}  
D_{Q_2}\left(\frac{z}{\zip}\right) \\ \label{boundd}
\end{eqnarray}
where $Q_0$, $\Qt_0$ are the transition scales at which the factorization scheme
is switched from $n_f$ to $n_f+1$, $n_f+2$ active flavors, respectively (assuming 
here for simplicity that $m_1<m_2$; a generalization to $m_1\ge m_2$ is 
straightforward). 
For general $Q_0$ also
the gluon density and $\alpha_s$ undergo a scheme
transformation. Here
we omit the corresponding formulae which can be found 
in \cite{ct,bmsn} since they are not needed in our analysis.
Canonically $Q_0$ is set equal to the heavy quark mass $m_1$ which guarantees 
\cite{ct} up to two loops a continuous evolution of $\alpha_s$ and the light
parton densities across $Q_0$. All available heavy quark densities 
are generated using $Q_0=m_1$ and we will therefore follow this choice here 
although a variation of $Q_0$ might substantially influence the heavy quark
results even far above the threshold \cite{grs}. Note that at three loops
a continuous evolution across $Q_0$ can no longer be achieved, neither for 
the parton distributions \cite{bmsn} nor for $\alpha_s$ \cite{alpsnnlo}.
Analogously to $Q_0=m_1$ we use $\Qt_0=m_2$ throughout. 
In Eq.\ (\ref{boundd}) we have made the distinction between the scale
dependent FF $D_{Q_2}(z,\Qt_0^2)$ and the scale independent FF $D_{Q_2}(z)$
explicit. Following the terminology in \cite{melnas,cagre} the latter
corresponds to the nonperturbative part of the former and describes
the hadronization process at the end of the parton shower which is
described perturbatively by the massless RG evolution. Alternatively, 
$D_{Q_2}(z)$ corresponds to a scale-independent FF within FOPT where no
collinear resummations are performed. These two points of view may
induce a scheme dependence if $D_{Q_2}(z)$ is fitted to data.
In principle, the massless evolution 
equations generate nonzero FF's also for light partons to decay into heavy
flavored hadrons. These light$\rightarrow$heavy contributions  
are important at LEP energies \cite{cagre} 
but can be safely neglected at the scales considered here\footnote{
We could therefore, in principle, restrict the evolution of the charm FF to the
nonsinglet sector.}
and we will assume $D_{i\neq Q_2}(z,\mu^2)=0$ throughout.

\subsection{SI Structure Functions at \alps1}

In the next section we will consider three types of \alps1 VFNS structure
functions. The first two are constructed at full \alps1
\begin{equation}
\label{full}
 F_i^{QS^{(0+1)}+GF} - F_i^{SUB_q+SUB_g+[SUB_D]}\ \ \ ,
\end{equation}
where the inclusion or omission of the bracketed $SUB_D$ term corresponds
to a running or scale-independent fragmentation function, respectively,  
as discussed in
the previous section. It is somewhat unclear whether $QS^{(1)}$ 
contributions (and the corresponding subtractions) 
should be considered on the same perturbative level as $GF^{(1)}$, see
\cite{kschie1} for a more detailed discussion on that point. In the original
formulation of the ACOT scheme \cite{acot} $QS^{(1)}$ contributions are neglected 
at the level we are considering here and we therefore do also consider this
option via the partial \alps1 structure function
\begin{equation}
\label{partial}
F_i^{QS^{(0)}+GF} - F_i^{SUB_g}\ \ \ .
\end{equation}   
For obtaining the numerical results of the next section the general formulae of 
this section have to be adjusted by chosing masses and couplings according 
to the relevant NC and CC values as listed in \cite{kschie1}.
For the CC case where $Q_1$ should be identified with strange quarks
the boundary condition in Eq.\ (\ref{boundq}) is inadequate
since $m_s^2\sim \Lambda_{QCD}^2$ is below the perturbative regime of
QCD. We will have recourse to standard strange seas from the literature 
\cite{grv94,cteq4} instead.

\section{The Charm Fragmentation Function in SI DIS} 

We will investigate the charm fragmentation function in CC and in NC SI DIS.
In CC DIS the charm production mechanism is undebated since charm is
dominantly produced by scattering on light strange quanta.
Our reasoning will therefore be that $D_c$ is directly accessible in 
CC DIS at relatively low spacelike momentum transfer. An extracted
$D_c$ can then be applied to NC DIS,
where it might give insight into the details of the 
production dynamics \cite{h1}. Also a test of the universality \cite{klre}
of the charm FF measured in CC DIS and $e^+e^-$ annihilation \cite{cagre} 
would be an important issue directly related to the factorization
theorems \cite{fact}
of perturbative QCD (pQCD). 

All $\varepsilon_c$ parameters 
discussed
below refer to a Peterson type \cite{peterson} functional form given by
\begin{equation}
\label{peterson}
D_c(z) = N \left\{ z \left[ 1-z^{-1}-\varepsilon_c/(1-z)
\right]^2\right\}^{-1}
\end{equation}
where $N$ normalizes $D_c$ to $\int dz D_c(z)=1$.

\subsection{CC DIS}

In CC DIS one does not expect to gain much insight into the charm
production process since 
charm is dominantly produced in scattering events on 
strange quarks\footnote{
We assume a vanishing Cabibbo angle. Our results remain, however, 
unchanged if the $CKM$ suppressed $d\rightarrow c$ background is included.
} 
in the nucleon, thereby permitting an experimental
determination of the strange quark content of the nucleon
\cite{cdhsw,ccfrlo,ccfrnlo}.
On the other hand the well understood production mechanism makes
a direct determination of the charm fragmentation function feasible
by measuring the energy spectrum of final state charm fragments. 
This is obvious in leading order accuracy where\footnote{
We will suppress some obvious scale-dependences in the following 
formulae and in their discussion.
} 
\begin{equation}
\label{simple}
d \sigma_{LO} \propto s(\chi) D_c(z) 
\end{equation}
is directly proportional to $D_c$.
More precisely, it is not $z$ but the closely related energy of the 
$c\rightarrow \mu\nu$ decay muon which can be observed in iron detectors
\cite{cdhsw,ccfrlo,ccfrnlo}. The smearing effects of the decay complicates the
determination of the precise shape of $D_c$ but only weakly influences an
extraction of $\left<z\right>$ \cite{cdhsw} which is valuable information 
if physically motivated one-parametric ans\"{a}tze \cite{peterson,colspi}
for $D_c$ are {\it{assumed}}.  
At NLO the production cross section is no longer of the simple factorized
form of Eq.\ (\ref{simple}) \cite{gkr2} and double convolutions 
(symbol $\otimes$ below) of the
form of Eqs.\ (\ref{QS1}), (\ref{GF}) have to be considered.
However, to a reasonable approximation
\begin{eqnarray} \nonumber
d \sigma_{NLO} &=& \left( \left[ s \otimes d{\hat{\sigma}}_s 
+ g \otimes d{\hat{\sigma}}_g \right ] \otimes D_c \right) 
(x,z,Q^2) \\ \nonumber
&\equiv& d \sigma_{LO}\ K(x,z,Q^2) \\ \nonumber
&\propto& s(\chi)\ D_c(z)\ K(x,z,Q^2) \\
&\simeq& s(\chi)\ {\cal{D}}_{x,Q^2}[D_c](z) 
\label{kfactor}
\end{eqnarray}
holds also at NLO accuracy within experimental errors
and for the limited kinematical range of present data
on neutrinoproduction of charm. In Eq.\ (\ref{kfactor}) the approximate 
multiplicative factor ${\cal{D}}$ absorbs the precise $K$-factor $K(x,z,Q^2)$
obtained from a full NLO QCD calculation \cite{gkr2}. ${\cal{D}}$
is {\it{not}} a simple universal fragmentation
function but a nontrivial process-dependent functional 
which is, however, mainly sensitive on $D_c$ and shows little sensitivity on
the exact parton distributions considered. 
The occurrence of $x,Q^2$ and $z$ in Eq.\ (\ref{kfactor}) as indices and 
as a functional argument, respectively, reflects the fact that the dependence
on $x$ and $Q^2$ is much weaker than is on $z$. 
Eq.\ (\ref{kfactor}) tells us that $s(\chi)$ fixes the normalization of
$d \sigma$ once $K$ is known. On the other hand $K$ (or ${\cal{D}}$) can be
computed \cite{gkr2} from $D_c$ with little sensitivity 
on $s(\chi)$, 
such that $s(\chi)$ and $D_c(z)$ decouple in the production
dynamics and can be simultaneously extracted. 
This point can be clearly inferred from Fig.\ 1 where it is shown
that the wide spread of CC charm production predictions which were obtained
in \cite{gkr2} using GRV94 \cite{grv94} and CTEQ4 \cite{cteq4} strange seas
can be brought into good agreement by a mere
change of the normalization given by the ratio 
$s_{GRV}(\chi)/s_{CTEQ4}(\chi)$.
The remaining difference is not within present experimental
accuracy which can be inferred from the shaded band representing
a parametrization \cite{gkr2} of CCFR data \cite{ccfrlo}.  
High statistics neutrino data therefore offer an ideal
scenario to measure $D_c$ complementary to an extraction
from LEP data on $e^+e^- \rightarrow D X$ \cite{cagre,bkk1,bkk2}.
This has been first noted in \cite{klre} where also a successful
test of the universality of the charm FF has been performed. 
With new data \cite{cdhsw,ccfrlo,ccfrnlo}\footnote{
Data from NuTeV \cite{nutev} is to be expected in the near future;
$\mu^\pm$ events observed at NOMAD await further analysis \cite{zuber}.
}
at hand and with 
a sounder theoretical understanding \cite{gkr2} of neutrinoproduction of
charm it would be desirable to update the analysis in \cite{klre}.
Nowadays one can in principle 
examine the possibility of a uniform renormalization group
transformation from spacelike momenta near above the charm mass 
($\nu N\rightarrow D X$) to
timelike momenta at the $Z^0$ peak ($e^+e^-\rightarrow D X$). 
In \cite{cagre,bkk1,bkk2} charm fragmentation functions extracted from
LEP data have been tested against $p_T$ and $\eta$ distributions
measured in photoproduction at HERA. 
However, we believe that a comparison to $z$ differential neutrinoproduction 
data is worthwhile beyond, since the latter measure directly 
the fragmentation spectrum
whereas $p_T$ and $\eta$ shapes are rather indirectly influenced by
the precise hardness of the FF via their 
normalization \cite{cagre,zeusgamma}.
Unfortunately up to now no real production data are available
but only strange sea extractions \cite{cdhsw,ccfrlo,ccfrnlo}
resulting from an analysis of the
former. We therefore strongly recommend that experiments publish
real production data such that the above outlined program can be executed
with rigour.    
Here we can only find a $\varepsilon_c$ parameter which
lies in the correct ball park and examine a few points which will
become relevant for an extraction of $D_c$ once data will become
available.

An outstanding question is the possible effect of a finite strange mass
on the full semi-inclusive charm production cross section \cite{kschie2}
including \alps1 quark scattering contributions.
By comparing the thick and the thin solid curve in Fig.\ 2 (a) it is
clear that the effect of a finite $m_s$ can be neglected even at low
scales and for a maximally reasonable value of $m_s=500$ MeV.
For the larger scale of Fig.\ 2 (b) the effect of chosing a finite
$m_s$ would be completely invisible.
A further question which might influence the extraction of a universal
FF from neutrinoproduction is the one of the scheme dependence in handling
final state quasi-collinear logarithms $\ln (Q^2/m_c^2)$. If these are
subtracted from the coefficient functions as discussed in section 
\ref{subterms}, the subtraction defines a running of the charm FF which 
becomes scale dependent\footnote{
Evolving the charm FF we adopt for consistency the evolution parameters 
$m_{c,b}$ and $\Lambda^{QCD}_{4,5}$ of the CTEQ4(M) \cite{cteq4}
parton distribution functions and we use $\Qt_0=m_c$.    
}
according to Eq.\ (\ref{boundd}).
In Fig.\ 2 we examine such resummation effects for 
CCFR kinematics \cite{ccfrlo}. We use the same Peterson FF with 
$\varepsilon_c=0.06$ \cite{chrin,gkr2} once for a fixed order calculation
\cite{gkr2} (solid lines) and once as the
nonperturbative part on the right hand side 
of the entire $c\rightarrow D$ FF on the left hand side of 
Eq.\ (\ref{boundd}) (dashed curves).
We note that towards intermediate scales around 
$Q^2\sim 20 {\rm{GeV}}^2$
one begins to see the softening effects of the resummation
which are enhanced as compared to FOPT. However,
as one  would expect at these scales, the resummation effects are 
moderate  and could be compensated by only a {\it{slight}} shift of the
$\varepsilon_c$ parameter which is therefore, within experimental accuracy, 
insensitive to scheme transformations. We note that -- as was already shown
in \cite{gkr2} --  according to Fig.\ 1 an  $\varepsilon_c$ of around $0.06$ 
which we took from 
an older analysis in \cite{chrin} seems to reproduce the measured spectrum
quite well. 
For $\left<E_\nu\right>=80{\rm{GeV}}$, $\left< Q^2\right>=20{\rm{GeV}}^2$
a value $\varepsilon_c \simeq 0.06$ gives an average $\left<z\right>\simeq0.6$
consistent with $\left<z\right>=0.68\pm 0.08$ measured by CDHSW \cite{cdhsw}.
In \cite{cagre}\footnote{
A similar $\varepsilon_c$ has been obtained in \cite{bkk1} in a related scheme.
The $\varepsilon_c$ value in \cite{bkk2} has no connection to a massive
calculation and cannot be compared to the values discussed here.
}
a distinctly harder
value of $\varepsilon_c \simeq 0.02$ was extracted from LEP data
on $e^+e^-\rightarrow D^\ast X$. 
If the latter fit is
evolved down to fixed target scales it is -- even within the
limited experimental accuracy -- incompatible with the  CCFR neutrino 
data represented
in Fig.\ 1. From Fig.\ 2 it is clear that the difference cannot be attributed to
a scheme dependence of the $\varepsilon_c$ parameter which is too
small to explain the discrepancy. It would of course be interesting to know
how much the above mentioned smearing effect of the $c\rightarrow \mu\nu$ decay
might dilute the discrepancy.   
In any case, charm fragmentation at LEP
has been measured by tagging on $D^\ast$'s whereas neutrinoproduction
experiments observe mainly $D$'s through their semileptonic decay-channel
(dimuon events). ARGUS \cite{argus} and CLEO \cite{cleo} data at 
$\sqrt{s}\simeq 10 {\rm{GeV}}$ indeed show \cite{pdg} 
a harder energy distribution of $D^\ast$'s compared
to $D$'s. It seems therefore to be possible within experimental accuracy to observe 
a nondegeneracy of the charm fragmentation functions into the lowest
charmed pseudoscalar and vector mesons.  
We note that an $\varepsilon_c$ value around $0.06$ which is in agreement
with neutrino data on $D$-production
is also compatible with the $D$ energy spectrum measured
at ARGUS where the evolution may be performed either via FOPT using 
expressions in \cite{nasweb} or via a RG transformation along the lines of
\cite{melnas,cagre}. If forthcoming experimental analyses should confirm our 
findings the lower decade $m_c(\sim 1{\rm{GeV}})\rightarrow$ ARGUS$(10{\rm{GeV}})$
may be added to 
the evolution path ARGUS$(10{\rm{GeV}})\rightarrow$LEP$(M_Z)$ paved in 
\cite{cagre} for the charm FF.

\subsection{NC DIS}

The fragmentation function of charm quarks observed in NC DIS is of special 
interest since it allows for directly investigating \cite{h1,zeus}
the charm production
mechanism which is a vividly discussed issue in pQCD 
phenomenology \cite{grs,bmsn,hqs}. 
Whereas for intrinsic heavy quarks one expects to observe a Peterson-like
hard spectrum attributed 
to the dominance of the leading order quark scattering
contribution, one expects a much softer spectrum for extrinsic GF since
the gluon radiates the $c\bar{c}$ pair towards lower energies ($z$) during
the hard production process before the nonperturbative hadronization takes 
place. 
Experimental analyses have been performed in \cite{h1,zeus} and the steep 
spectrum\footnote{
The variable $x_D$ considered in \cite{h1,zeus} differs slightly from $z$ in 
definition. 
The variables are, however, identical at the 2\% level \cite{daum}.
} (best visible in Fig.\ 6 in \cite{h1})
of observed $D$-mesons together with the missing of a hard 
component at larger $z$ seem to give clear evidence for the dominance of 
extrinsic $GF$ over $QS$ which was excluded at the 5\% level \cite{h1}.
In a complete VFNS the $QS^{(0)}$ component makes, however, just one part 
of the
\alps1 structure functions in Eqs.\ (\ref{full}), (\ref{partial}). 
Especially, there also exists a $GF^{(1)}$ component, albeit with 
the leading log part of it subtracted. 
Since the subtraction term in Eq.\ (\ref{SUBg}) is proportional to  
$D_c$ and therefore only removes a hard component from $GF^{(1)}$ one expects 
the rise towards lower $z$ to survive the subtraction. Furthermore a 
perturbative evolution of the
charm fragmentation function $D_c(z,Q^2)$ might soften somewhat the
hard QS term.

These expectations can be quantitatively confirmed in Fig.\ 3 where
we show for HERA kinematics \cite{h1}
the total (solid line) normalized \alps1 production cross section for 
transverse virtual 
photons ($F_L=0$)
on protons. We also show the individual components contributing to it:
The processes $\gamma^\ast g\rightarrow c {\bar{c}}$ 
(dot-dashed) and
$\gamma^\ast c \rightarrow c g$ 
(dotted; incl.\ virtual corrections) correspond
to the $GF^{(1)}-SUB_g$ and $QS^{(1)}-SUB_q-SUB_D$ terms, respectively,  
subtracted at $\mu=Q$. They
are {\it{not}} physically observable and only sensible if they are
added to the $QS^{(0)}$ Born term (dashed) as in Eqs.\ (\ref{full}), 
(\ref{partial}). 
We have perturbatively resummed all logarithms of the charm mass via massless 
evolution equations starting at the charm mass [$Q_0=\Qt_0=m_c$]
and using the standard boundary 
conditions in Eqs.\ (\ref{boundq}), (\ref{boundd}) for $\varepsilon_c=0.06$. 
Finite charm mass effects on the subtracted $QS^{(1)}$ contribution can be
inferred by comparing the thick and the thin dotted curves, where the 
$m_c \rightarrow 0$ limit has been taken for the latter. 
As has been
theoretically anticipated in \cite{collins} the charm mass can be safely set
to zero in $QS^{(1)}$ and the involved convolutions in Eq.\ (\ref{ansatz})
may be replaced by the massless expressions in \cite{aemp,fupe} which simplifies 
the numerics essentially and which we will therefore do for the $\mu=2 m_c$ curve
in Fig.\ 4 below.
As also stressed in \cite{collins} it is, however, essential to keep
the charm mass finite in the $GF^{(1)}$ contribution since $m_c$ tempers
the strength of the ${z^\prime}^{-1}$ ${\hat{u}}$-channel 
propagator singularity.
Obviously the \alps1 result of an ACOT based calculation deviates essentially
from the naive Born term expectation and it seems by no means legitimate
to treat $GF^{(1)}$ as a higher order correction here. Contrary to the
corresponding inclusive results in \cite{kschie1} and to the expectations in 
\cite{acot} also the subtracted $QS^{(1)}$ term is numerically significant
in the semi-inclusive case considered here. In the light of the huge 
\alps1 corrections it seems, however, undecidable as to whether include
or omit $QS^{(1)}$ at \alps1 without knowing \alpsq\ corrections within ACOT.

In Fig.\ 4 the resummed result (VFNS: solid lines)
can be compared to 
unsubtracted \alps1\footnote{
An ${\cal{O}}(\alpha_s^2)$ NLO calculation \cite{hasm} 
within FOPT gives very similar results \cite{daum}. 
} 
$GF^{(1)}$ (fixed order, dashed line). 
We show the total \alps1 VFNS result for the choices
$\mu=Q,2 m_c$. 
The modest  scale dependence arises exclusively through the $QS^{(1)}$ term.
For any of the other curves a variation of $\mu$ is completely insignificant and we
therefore only show $\mu=Q$.
A full VFNS calculation (solid lines)
seems hardly distinguishable from
fixed order perturbation theory (dashed line) within experimental accuracy. The two
approaches are even closer if one follows the suggestion in \cite{acot} 
and does not include 
(dot-dashed line) the subtracted $QS^{(1)}$ term at the level of 
$QS^{(0)}+GF^{(1)}$. 
The data points in the figure correspond to H1 measurements \cite{h1}
of $D^{0}$ (circles) and ${D^{\ast}}^+$ 
(triangles; both including charge conjugation) spectra.
The measurement is restricted to $\eta_D<1.5$. Extrapolating to the full
phase space gives rise to large acceptance corrections which are, however,
quite uniform \cite{daum} over the kinematical range considered and therefore
have a minor effect on the {\it{normalized}} spectrum.   
Since FOPT and ACOT based calculations are very close it seems improbable
that an experimental discrimination between the two approaches
will be possible. The born term in Fig.\ 3 is far from being the dominant 
contribution and an intrinsic $c(x,Q^2)$ stemming from the resummation of
perturbative logs can therefore not be excluded.
The tendency of the data appears somewhat softer than any of the 
calculations and the tendency seems to be confirmed by preliminary
data in a lower $z$ bin \cite{tza}. The resummed calculation appears to be 
too hard at larger $z$ around $0.6$ if the $QS^{(1)}$ component is included. 
As already mentioned,
at the present stage of the calculations it can not be decided whether this
hints at an intrinsic problem within VFNS calculations or wether this may
be cured by \alpsq\ corrections.

\section{Conclusions}
In this paper we have performed an ACOT \cite{acot} based analysis of heavy quark 
fragmentation in DIS including a calculation of semi-inclusive scattering on 
massive quarks at \alps1 for general masses and couplings. 
As in the inclusive case \cite{kschie1} effects from finite initial state
quark masses can be neglected for practical applications to charm
production in CC and NC DIS. The involved convolutions in section 
\ref{scattq} can therefore safely be replaced by their analogues in
\cite{aemp,fupe,gkr2}.  
Neutrinoproduction is an ideal environment to extract the charm FF
within DIS and a Peterson \cite{peterson} type FF with 
$\varepsilon_c\simeq 0.06$ seems to lie in the correct ball park, where the 
sensitivity on the choice of scheme is small and finite $m_s$ effects
are irrelevant. 
The $\varepsilon_c$ value above is compatible with $e^+e^-$ data 
if a nondegeneracy of charm quarks fragmenting into $D$'s and $D^\ast$'s is
allowed for. 
For NC DIS it seems unlikely that a discrimination between
fixed order and resummed calculations will be possible at HERA. Both
approaches give similar results which show a spectrum that is somewhat harder
than the tendency of the data \cite{h1,zeus,tza}. The resummed calculation is
made worse if the \alps1 quark scattering contribution is included at the
perturbative level considered in this paper.     

\section*{Acknowledgements}

We thank E.\ Reya for advice, useful discussions and a careful reading of
the manuscript and K.\ Daum for
several helpful correspondences on the HERA analyses. 
This work has been supported in part by the
'Bundesministerium f\"{u}r Bildung, Wissenschaft, Forschung und
Technologie', Bonn.   

\newpage
\setcounter{equation}{0}
\def\theequation{A\arabic{equation}}
\section*{Appendix A:\\ 
Matrix Elements for Real Gluon Emission}
The projections $\hat{f}_i^Q$ of the partonic Matrix Element onto the structure functions
are most conveniently given in the Mandelstam variables $\hsi$ and $\hti$
which are defined below Eq.\ (\ref{Hiq}):
\begin{eqnarray} \nonumber
\hat{f}_1^Q(\hsi,\hti)&=&
\frac{8}{{\Delta^\prime}^2}\Bigg\{
 -\Delta^2 (\Sp \spp -2 m_1 m_2 \Sm)  
\left(\frac{m_2^2}{\hsi^2}+\frac{m_1^2}{\hti^2}+\frac{\spp}{\hsi \hti}\right)  
\\ \nonumber
&+&  2 m_1 m_2 \Sm \Bigg(
\frac{m_1^2 \hsi (\hsi +2\spm)}{\hti^2}
+ \frac{{\Delta^\prime}^2 +(m_2^2+Q^2)\hsi + 
2 \spm \spp}{\hti}
\\ \nonumber
&+& \frac{{\Delta^\prime}^2 -\hsi (m_1^2+Q^2 +\hsi)+2 m_2^2 \spm}{\hsi} 
- \hti \frac{(m_2^2+\hsi)}{\hsi} \Bigg)
\\ \nonumber
&+& \Sp \Bigg(
- \frac{{m_1}^2 \hsi \spm (\hsi+2 \spp)}{\hti^2}
+ \frac{-\hsi^3 -4\hsi^2 \spm + \hsi (4 m_1^2 m_2^2- 7\spm \spp)}{2 \hti}
\\ \nonumber
&+& \frac{2 \spp (-\Delta^2 - 2 \spm \spp)}{2 \hti}     
+\Bigg[4 m_1^4+ 2 m_1^2 \hsi - \spm (m_2^2 + \spm) 
\\ \nonumber
&-& 
\frac{(\Delta^2 + 2 m_2^2 \spm)  \spp}{\hsi}\Bigg]- \hti \frac{{\Delta^\prime}^2 
- 2 (m_2^2+\hsi) \spp}{2 \hsi}
\Bigg) \Bigg\}
\\ \nonumber
\hat{f}_2^Q(\hsi,\hti)&=&
\frac{16}{{\Delta^\prime}^4}\Bigg\{
 -2 \Delta^4 \Sp 
\left(\frac{m_2^2}{\hsi^2}+\frac{m_1^2}{\hti^2}+\frac{\spp}{\hsi \hti}\right)  
+ 2 m_1 m_2 \Sm 
\Bigg(\frac{(\delp^2-6 m_1^2 Q^2)\hsi}{\hti}
\\ \nonumber
&+& \left[2(\delp^2-3 Q^2(\hsi+\spp))\right]
+ \hti \frac{\delp^2-6 Q^2(m_2^2+\hsi)}{\hsi} \Bigg)
\\ \nonumber
&+& \Sp \Bigg(\frac{-2 m_1^2 \hsi [(\Delta^2-6 m_1^2 Q^2)\hsi
+2 \Delta^2 \spm]}{\hti^2}
+\frac{-2 \Delta^2(\Delta^2+2\spm\spp)}{\hti}
\\ \nonumber
&+& \frac{- \hsi[2(\Delta^2-6 m_1^2 Q^2)\hsi
+(\delp^2-18 m_1^2 Q^2)\spp+2 \Delta^2(3\spp-4 m_1^2)]}{\hti}
\\ \nonumber
&+& \Big[-2(m_1^2+m_2^2)\hsi^2-9 m_2^2\spm^2-\frac{2 \Delta^2(\Delta^2+2 m_2^2\spm)}{\hsi}
+ 2\hsi[2\Delta^2
\\ \nonumber
&+& (m_1^2- 5 m_2^2)\spm] + \Delta^2(2\spp-m_2^2)\Big]
-\hti \frac{[\delp^2-6 Q^2(m_2^2+\hsi)]\spp}{\hsi} 
\Bigg) \Bigg\}
\\ \nonumber
\hat{f}_3^Q(\hsi,\hti)&=&
\frac{16}{{\Delta^\prime}^2}\Bigg\{
 -2 \Delta^2 \Rp 
\left(\frac{m_2^2}{\hsi^2}+\frac{m_1^2}{\hti^2}+\frac{\spp}{\hsi \hti}\right)  
+ 2 m_1 m_2 \Rm 
\Bigg(\frac{\hsi+\spm}{\hti}+\frac{\hsi-\smp}{\hsi}\Bigg)
\\ \nonumber
&+& \Rp \Bigg(
\frac{- 2 m_1^2 \hsi \spm}{\hti^2}
+\frac{-\hsi^2-4 (\Delta^2- m_1^2 \smp) -3 \hsi \spm}{\hti}
\\ 
&+& \left[2(m_1^2-m_2^2)-\frac{2(\Delta^2+m_2^2 \spm)}{\hsi} \right] 
+\hti \frac{\hsi-\smp}{\hsi}
\Bigg) \Bigg\}
\end{eqnarray}
where we conveniently use
the shorthands $\Delta \equiv \Delta[m_1^2,m_2^2,-Q^2]$ 
and $\Delta^\prime \equiv \Delta[m_1^2,\hat{s},-Q^2]$. 
In order to obtain the inclusive results $\hat{f}_i^Q(\hsi)$ in
Appendix C of \cite{kschie1} the 
$\hat{f}_i^Q(\hsi,\hti)$ have to be integrated over $0\le y \le 1$, i.e.
\begin{equation}
\hat{f}_i^Q(\hsi)= \int_0^1 dy\ \hat{f}_i^Q(\hsi,\hti)\ \ \ ,
\end{equation}
where $y$ is defined via the the partonic center of 
mass scattering angle $\theta^\ast$ and related to $\hti$ through
\begin{eqnarray} \nonumber
y &\equiv& \frac{1}{2}\ (1+\cos \theta^\ast) \\
\hti&=&
\frac{\hsi}{\hsi+m_2^2}\ \delp\ (y-y_0)\ \ \ ,
\end{eqnarray}
with $y_0=[1+(\spp+\hsi)/\delp]/2$ being the would-be collinear pole of the
${\hat{t}}$-channel propagator.

\newpage
\section*{Figure Captions}
\begin{description}
\item[Fig.\ 1] 
The charm production cross section obtained in \cite{gkr2} for two $x,Q^2$ 
points in the CCFR \cite{ccfrlo} kinematical regime. Up to a constant of 
normalization which was
conveniently chosen in Eq.\ (4) of \cite{gkr2} $s_{eff}$ represents the triple 
differential cross section $d^3 \sigma / dx dy dz$ where $x$ and $y$ are
standard and $z\equiv p_D\cdot p_N/q\cdot p_N$. Shown are the predictions
using GRV94 \cite{grv94} (solid) and CTEQ4 \cite{cteq4} (dashed) partons and 
a curve (dot-dashed) where the normalization of the CTEQ4 prediction 
is changed by multiplying with
the ratio of the strange seas $s_{GRV}(\chi)/s_{CTEQ}(\chi)$. For all curves
a scale-independent Peterson FF with $\varepsilon_c=0.06$ has been used.
  
\item[Fig.\ 2] The same quantity as in Fig.\ 1. All predictions have been obtained
for the CTEQ4 parton distributions. The solid curves result from
a scale-independent Peterson FF with $\varepsilon_c=0.06$. In (a) 
the thin solid curve has
been obtained using the formulae of section \ref{sec2} and chosing 
a finite strange quark mass of $m_s=500 {\rm{MeV}}$ whereas
the thicker curve corresponds to the asymptotic $m_s \rightarrow 0$ limit
($\equiv$\msbar). In (b) these two options would be completely
indistinguishable and we only show the \msbar\ result. For the dot-dashed curves
the final state quasi-collinear logarithm has been absorbed into a running of
the charm FF. A Peterson FF with $\varepsilon_c=0.06$ has been used as the
nonperturbative part of the input as given in Eq.\ (\ref{boundd}).

\item[Fig.\ 3] The normalized charm production cross section 
$d \sigma / dz$ ($z\equiv p_D\cdot p_N/q\cdot p_N$) for HERA kinematics
($\sqrt{s}=314 {\rm{GeV}}$). 
For a comparison with H1 data \cite{h1} the cross 
section has been integrated over $10 {\rm{GeV}}^2< Q^2 < 100 {\rm{GeV}}^2$
and $0.01<y<0.7$. Following the experimental analysis \cite{h1} only
contributions from transverse photons ($F_L=0$) are considered.   
Shown is the total \alps1 result (solid line) and the individual contributions.
Details to the calculation of the total result and the individual contributions
are given in the text. The charm mass has been kept finite at the
$CTEQ4$ value of $m_c=1.6 {\rm{GeV}}$ everywhere except for the thin dotted
curve where the $m_c\rightarrow 0$ limit has been taken.
 
\item[Fig.\ 4] A Comparison of the total \alps1 result (solid lines)
to H1 data \cite{h1} on $D^0$ (circles) and ${D^+}^\ast$ (triangles; both including
charge conjugation) production. Shown are results for two choices of the
factorization/renormalization scale $\mu$. 
Also shown is the outcome of a fixed order \alps1
$GF$ calculation for comparison (dashed line). The dot-dashed line follows
the suggestion in \cite{acot} and neglects quark initiated contributions
at \alps1, i.e. the difference between the solid ($\mu=Q$)
and the dot-dashed line is given
by the (thick) dotted line in Fig.\ 3. For the dot-dashed curve as well as for the 
fixed order calculation (dashed)
only $\mu=Q$ is shown since the scale dependence is
completely insignificant.    

\end{description}

\newpage
\pagestyle{empty}
\begin{figure}
\vspace*{-1cm}
\hspace*{-1.5cm}
\epsfig{figure=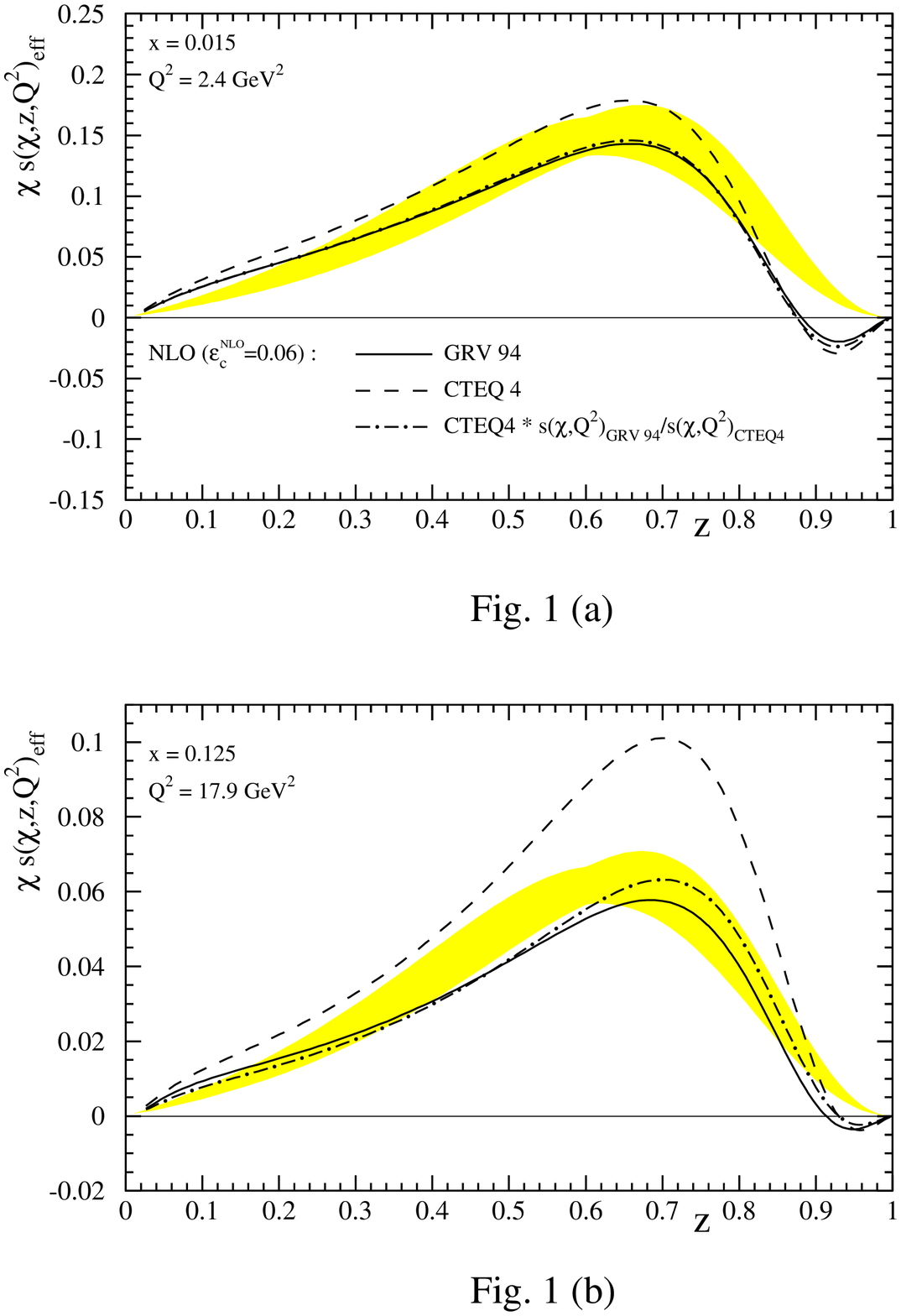,width=20cm}
\end{figure}
\newpage
\begin{figure}
\vspace*{-1.5cm}
\hspace*{-2.5cm}
\epsfig{figure=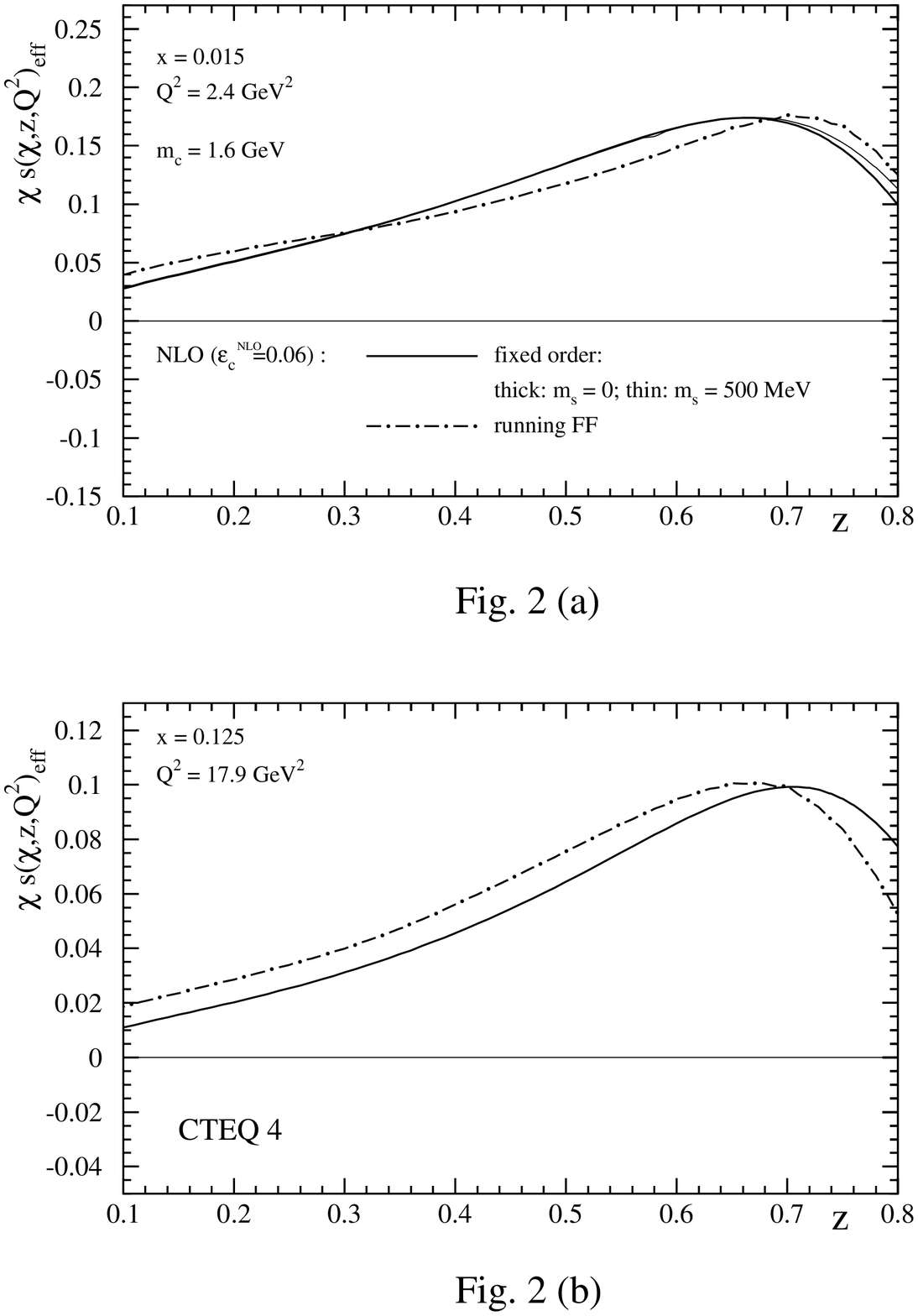,width=20cm}
\end{figure}
\newpage
\begin{figure}
\vspace*{-1.5cm}
\hspace*{-2.5cm}
\epsfig{figure=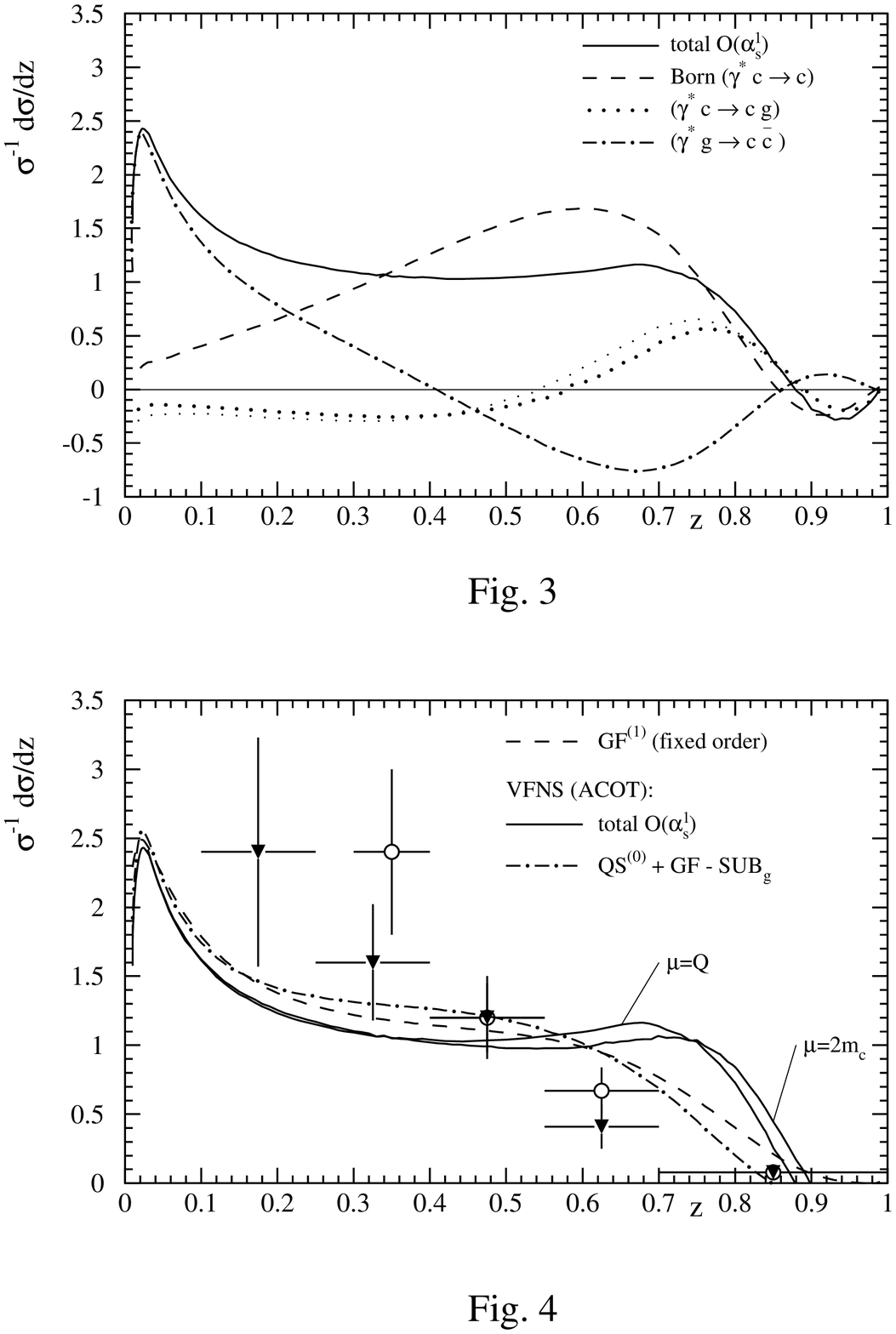,width=20cm}
\end{figure}
\end{document}